\documentclass[journal]{IEEEtran}

\ifCLASSINFOpdf
\else
   \usepackage[dvips]{graphicx}
\fi
\usepackage{url}

\usepackage{graphicx}
\usepackage{booktabs}
\usepackage{multicol}
\usepackage{multirow}
\usepackage{subfigure}
\usepackage{amsmath}
\usepackage{amssymb,color}
\usepackage{ulem}

\begin{document}

\title{Diffusion-Based Adversarial Purification for Speaker Verification}

\author{Yibo Bai, Xiao-Lei Zhang, and Xuelong Li
\thanks{Corresponding author: Xiao-Lei Zhang}
\thanks{Yibo Bai is with the Department of Electrical and Electronic Engineering, The University of Hong Kong, Hong Kong (e-mail: baiyibo@connect.hku.hk).}
\thanks{Xiao-Lei Zhang is with the School of Marine Science and Technology, Northwestern Polytechnical University, 127 Youyi West Road, Xi'an, Shaanxi 710072, China, with the Institute of Artificial Intelligence (TeleAI), China Telecom Corp Ltd, 31 Jinrong Street, Beijing 100033, P. R. China, and also with the Research and Development Institute of Northwestern Polytechnical University, Shenzhen 518063, China. (e-mail: xiaolei.zhang@nwpu.edu.cn).}
\thanks{Xuelong Li is with TeleAI, China Telecom Corp Ltd, 31 Jinrong Street, Beijing 100033, P. R. China (e-mail: xuelong\_li@ieee.org).}
}

\markboth{Journal of \LaTeX\ Class Files, Vol. 14, No. 8, August 2015}
{Shell \MakeLowercase{\textit{et al.}}: Bare Demo of IEEEtran.cls for IEEE Journals}
\maketitle

\begin{abstract}
Recently, automatic speaker verification (ASV) based on deep learning is easily contaminated by adversarial attacks, which is a new type of attack that injects imperceptible perturbations to audio signals so as to make ASV produce wrong decisions. This poses a significant threat to the security and reliability of ASV systems. To address this issue, we propose a Diffusion-Based Adversarial Purification (DAP) method that enhances the robustness of ASV systems against such adversarial attacks. Our method leverages a conditional denoising diffusion probabilistic model to effectively purify the adversarial examples and mitigate the impact of perturbations. DAP first introduces controlled noise into adversarial examples, and then performs a reverse denoising process to reconstruct clean audio. Experimental results demonstrate the efficacy of the proposed DAP in enhancing the security of ASV and meanwhile minimizing the distortion of the purified audio signals.
\end{abstract}


\IEEEpeerreviewmaketitle

\section{Introduction}

\IEEEPARstart{A}{utomatic} speaker verification (ASV) aims to verify individuals based on their unique voiceprint characteristics. It has been widely used in biometric authentication. However, ASV systems are vulnerable to attackers \cite{villalba2020x}, which raises a concern in enhancing their security in real-world applications. Particularly, recently a new type of attacker named \textit{adversarial attack}, which adds imperceptible perturbations to original utterances, can easily contaminate an ASV system by making it e.g. accept speakers that should have been rejected or just the opposite, without changing the perception quality of the utterances to human. The polluted utterances are called \textit{adversarial examples}, while the original utterances are also called \textit{genuine examples}.
In recent years, there has been growing interest in studying the susceptibility of ASV systems to adversarial attacks. For example, \cite{villalba2020x,li2020adversarial} found that the state-of-the-art ASV models are highly vulnerable to adversarial attacks. \cite{kreuk2018fooling} conducted transferable gray-box attacks on ASV systems across different features and different models.

In response to this emerging threat landscape, researchers have begun investigating techniques to enhance the robustness and security of ASV systems against adversarial attacks. Current defense methods can be classified into three categories: adversarial training, adversarial detection and adversarial purification \cite{wu2023defender}. Specifically, adversarial training mainly utilizes the adversarial examples to retrain the ASV model. One weakness of this kind of methods is that it needs to modify the parameters of the original model \cite{goodfellow2014explaining}. Adversarial detection adds a detection head in front of the ASV model to reject adversarial examples as an input into the system \cite{li2020investigating}. However, it may hinder human access to ASV when his/her voice was polluted by adversarial perturbations. Adversarial purification aims to purify all incoming inputs to eliminate adversarial perturbations \cite{wu2023defender}. It overcomes the weaknesses of the first two kinds of methods, which is our focus in this paper.

Adversarial purification for ASV can be divided into two categories: preprocessing and reconstruction. Preprocessing methods apply empirical knowledge to the input signals. They are typically data-free and have low computational complexity. For example, \cite{wu2020defense} applied median, mean and Gaussian filters to the input utterance. \cite{chang2021defending} proposed to add white noise with different variance to the entire utterance. Reconstruction methods focus on recover the original audio or its acoustic features from the adversarial examples. \cite{zhang2020adversarial} proposed a separation network to estimate adversarial noise for restoring the clean speech. \cite{wu2021improving} proposed to reproduce the acoustic features with a self-supervised model. Although existing purification methods have demonstrated effectiveness in defending ASV systems, the quality of the reconstructed audio signals was not guaranteed to high level. Some methods introduce additional noise into the purified samples, while others produce unexpected distortion to the audio signals which make the signals deviate significantly from their origins.

To address the above issue, we propose the Diffusion-Based Adversarial Purification (DAP) method. This novel method utilizes a diffusion model to purify the impact of adversarial attacks by reconstructing the original speech waveform, which defends ASV systems with high reliability. Our contributions can be summarized as follows: We propose the first adversarial defense diffusion model for ASV systems. Our method achieves the state-of-the-art performance on the ASV purification task, and retains the essential information of the original speech signal.

\begin{figure*}[t]
\begin{center}
\includegraphics[width=0.56\textwidth]{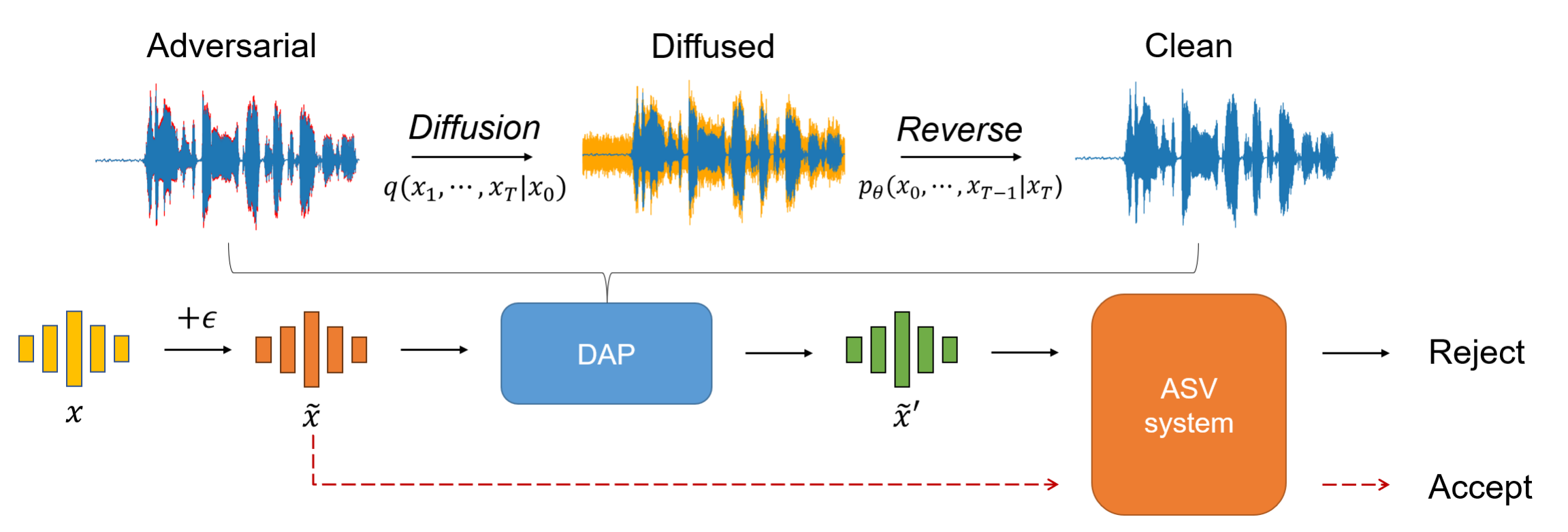}
\end{center}
\caption{A speaker verification pipeline with DAP method. Initially, the adversarial example is introduced into a diffusion model positioned before the ASV system for processing. Subsequently, the diffusion model employs a "diffusion" process on the adversarial input, followed by the reversal of this process to reconstruct the original clean audio. Finally, the ASV system produces the correct verification outcome.}
\label{pipeline}
\end{figure*}

\section{Related work}

\subsection{Automatic speaker verification}
Speaker verification aims to determine whether a test utterance belongs to a speaker that it declares to. Most of the current ASV systems comprise three components: an acoustic feature extractor, an encoder front-end which yields a speaker embedding from the acoustic features, and a scoring back-end which evaluates the similarity of the representations of two utterances. Commonly used acoustic features include Mel-frequency cepstral coefficients (MFCCs) or logarithmic filter-banks (LogFBank). Given a test utterance $x^t$ and an enrollment utterance $x^e$, the scoring process of ASV can be defined as:
\begin{align}
  s = S(F(x^t), F(x^e))
\end{align}
where $S(\cdot)$ denotes the scoring back-end, $F(\cdot)$ is the encoder front-end, and $s$ is the similarity score between $x^t$ and $x^e$. By comparing the similarity score with a predefined threshold, the system determines whether to accept the test utterance.

\subsection{Adversarial attack to ASV}

Given a genuine audio utterance $x$ from a speaker $i$, an adversarial attack creates a perturbation signal $\epsilon$ to $x$. The adversarial example is formulated as $\tilde{x}=x+\epsilon$ subject to the condition $\left\Vert
                    {\tilde{x} - x}
                    \right\Vert_p \leq\varepsilon$ which guarantees that $\tilde{x}$ is similar to $x$, where $\varepsilon$ is a very small number that controls the energy of $\epsilon$, and $\|\cdot\|_p$ is the $\ell_p$-norm. As shown in Fig. \ref{pipeline}, $\tilde{x} $ aims to cause an error of the ASV system.

\subsection{Denoising diffusion probabilistic models}

The denoising diffusion probabilistic models \cite{ho2020denoising} are a type of generative models used to produce data similar to the input. Specifically, the diffusion model works by progressively adding Gaussian noise to blur the training data and then learning how to denoise it and recover the original input. Once trained, the diffusion model can reverse the diffusion process to generate new data from random noise.

Recently, diffusion models have garnered interest among researchers. They utilize diffusion and denoising processes for high-quality content generation. By incorporating specific generation conditions, the outcomes of diffusion models can be controlled \cite{choi2021ilvr} to satisfy different applications such as speech enhancement, speech command recognition, image reconstruction, and remote sensing \cite{lu2022conditional,yu2023universal,wu2023defending,alkhouri2023diffusion}.

\section{Methodology}

\subsection{Framework}

The objective of our research is to develop a robust adversarial purification model $D(\cdot)$ for ASV, which eliminates the perturbation $\epsilon$ in $\tilde{x} $ and produces a purified audio $\tilde{x} ^{\prime}$. It can be formulated as a problem of $\tilde{x}^{\prime}=D(\tilde{x})$ subject to:
\begin{align}\label{score}
S(\tilde{x}^{\prime}, x^e) = S(x, x^e)
\end{align}

This paper proposes to take the denoising diffusion probabilistic model \cite{ho2020denoising} as $D(\cdot)$ to purify adversarial perturbation and transform these adversarial examples into clean data for ASV. The proposed DAP method is illustrated in Fig. \ref{pipeline}. Given an adversarial audio $\tilde{x}$, DAP purifies it to $\tilde{x}'$ for satisfying Eq. \eqref{score}. Specifically, DAP first introduces noise to $\tilde{x}$ via a \textit{forward process} with a diffusion timestep $T$. Subsequently, it reconstructs the clean audio signal $\tilde{x}^{\prime}$ via a reverse denoising process before feeding it into the ASV system for analysis. In the next subsection, we will present the denoising diffusion probabilistic model in detail.

\subsection{Diffusion-Based audio purification}

  A $T$-step denoising diffusion probabilistic model consists of two processes: the diffusion process, a.k.a. \textit{forward process}, and the reverse process, which can both be represented as a $T$-step parameterized Markov chain. In its diffusion process, the model adds $T$ rounds of noise to the real example $x_0$ to obtain the noised sample $x_T$. The reverse process aims to recover the original $x_0$ based on  $x_T$. Following the Markov assumption, the state at $t$ step in the diffusion process, only depends on the state at $t-1$ step, so the process can be defined as:
\begin{align}
  q(x_1,\cdots,x_T|x_0) = \prod_{t=1}^{T} q(x_t|x_{t-1}),
  \label{forward}
\end{align}
where $q(x_t|x_{t-1})=\mathcal{N}(x_t;$ $\sqrt{1-\beta_t}x_{t-1},\beta_t\mathbf{I})$. In other words, $x_t$ is sampled from a Gaussian distribution with mean $\sqrt{1-\beta_t}x_{t-1}$ and variance $\beta_t$, where $\beta_{t}$ is a hyperparameter determined by a predefined strategy, usually satisfying $\beta_1<\beta_2<\cdots<\beta_T$.

Then, employing the recursive reparameterization trick, $x_t$ can be represented in terms of $x_0$:
\begin{equation}
  x_t = \sqrt{\bar{\alpha}_t}x_0+(1-\bar{\alpha}_t)\boldsymbol{\epsilon},
\label{x_t}
\end{equation}
where $\boldsymbol{\epsilon}\sim\mathcal{N}(0,\mathbf{I})$, $\alpha_t = 1-\beta_t$ and $\bar{\alpha}_t =  \prod_{i=1}^t \alpha_i$. We have:
\begin{equation}
  q(x_t|x_0) = \mathcal{N}(x_t;\sqrt{\bar{\alpha}_t}x_0,(1-\bar{\alpha}_t)\mathbf{I}).
\end{equation}
Similarly, for the reverse process that transforms $x_T$ back to $x_0$, we have:
\begin{equation}
  p_\theta(x_0,\cdots,x_{T-1}|x_T) = \prod_{t=1}^T p_\theta(x_{t-1}|x_t),
  \label{reverse}
\end{equation}
where $p_\theta(\cdot)$ is used to estimate $q(\cdot)$ in Eq. \eqref{forward} and $p_\theta(x_{t-1}|x_t)=\mathcal{N}(x_{t-1};\mu_\theta(x_t,t),\sigma_\theta(x_t,t)^2\mathbf{I})$ with the parameterized $\mu_\theta$ and $\sigma_\theta^2$ described as \cite{kong2020diffwave}:
\begin{equation}
  \mu_\theta(x_t,t)=\frac{1}{\sqrt{\alpha_t}}(x_t-\frac{\beta_t}{\sqrt{1-\bar{\alpha}_t}}\boldsymbol{\epsilon}_\theta(x_t,t)),
 \end{equation}
and
\begin{equation}
  \sigma_\theta(x_t,t)^2=\tilde{\beta}_t,
 \end{equation}
where $\tilde{\beta}_t = \frac{1-\bar{\alpha}_{t-1}}{1-\bar{\alpha}_t}\beta_t$ for  $t>1$ and $\tilde{\beta}_1=\beta_1$, and $\boldsymbol{\epsilon}_\theta(x_t,t)$ represents a deep neural network model used to predict Gaussian noise $\boldsymbol{\epsilon}$ from $x_t$ and $t$.

According to \cite{ho2020denoising}, we train diffusion model with the following unweighted objective function:
\begin{equation}
  \mathbb{E}_{x_0,t,\epsilon}
                    \left\Vert
                    \boldsymbol{\epsilon} - \boldsymbol{\epsilon}_\theta(\sqrt{\bar{\alpha}_t} x_0 + \sqrt{(1 - \bar{\alpha}_t)} \boldsymbol{\epsilon}, t)
                    \right\Vert_2^2,
 \end{equation}
where $t$ is uniformly sampled from the range 1 to $T$. After training, $\boldsymbol{\epsilon}_\theta$ is able to predict $\boldsymbol{\epsilon}$ well. In the inference stage, the diffusion model first begins with a diffused sample $x_T$, then the reverse process iteratively utilizes $\boldsymbol{\epsilon}_\theta$ to get the mean $\mu_\theta$ and finally recover $x_0$ from $x_T$.

Existing theorems have proven that in the forward process Eq. \eqref{forward} of the diffusion model, the KL divergence between the distribution of clean data and the distribution of adversarial examples monotonically decreases \cite{nie2022diffusion}. This indicates that the two distributions gradually become more similar as $t$ increases, enabling the use of the reverse process to reconstruct clean inputs from adversarial examples. This comprehensive process constitutes the foundation of our innovative defense approach against adversarial attacks. While the previous method in \cite{nie2022diffusion} focuses on purifying images with fixed width and height, our method can flexibly handle audio signals with variable length. Given an adversarial audio $\tilde{x}$, we initiate the forward process with $x_0=\tilde{x}$. According to the above theorem in \cite{nie2022diffusion}, there exists a timestep $t^*$ that minimizes the KL divergence between the two distributions. Therefore, by starting the reverse process with $T=t^*$, the diffusion model can recover corresponding clean audio of $\tilde{x}$. The resulting audio utterance is then passed to the ASV system.

\section{Experiments}

\subsection{Experimental settings}

\subsubsection{Dataset}

{We utilized VoxCeleb1 \cite{nagrani2017voxceleb} for evaluating our DAP approach and VoxCeleb2 \cite{chung2018voxceleb2} for training the ASV and DAP models. VoxCeleb1 contains over 1,000 hours of speech data with 148,642 utterances from 1,211 speakers. VoxCeleb2 extends VoxCeleb1 with more speakers. This dataset offers an expanded set of development data to train our ASV models, which enables them to learn from a more diverse set of speakers. To balance computational requirements and ensure a representative evaluation, we conducted our adversarial research by randomly selecting 1,000 trials from the VoxCeleb1-O subset.}

\begin{table}[t]
  \centering
  \caption{ECAPA-TDNN performance results for genuine examples on the speaker verification task.}
  \setlength{\tabcolsep}{7mm}{
    \scalebox{1}{
    \begin{tabular}{ccc}
    \toprule
    Trials&EER(\%)&minDCF\\
    \midrule
    VoxCeleb-O&1.069&0.107\\
    VoxCeleb-E&1.201&0.131\\
    VoxCeleb-H&2.288&0.226\\
    1,000 trials&0.828&0.021\\
    \bottomrule
    \end{tabular}
    }
    }
  \label{tab:asv}
 \vspace{-2em}
\end{table}

\begin{table*}[t]
  \centering
  \caption{EER(\%) results of the victim ASV model for genuine and adversarial examples, given the defense models of TERA, spatial smoothing, adding noise and the proposed DAP method. The term ``N/A'' means that no defense model is applied.}
  \vspace{-0.5em}
  \scalebox{0.95}{
    \begin{tabular}{cccccccccccccc}
    \toprule
     & \multirow{2}[0]{*}{N/A} &\multicolumn{2}{c}{TERA \cite{wu2021improving}}&\multicolumn{3}{c}{Spatial smoothing \cite{wu2020defense}}&\multicolumn{5}{c}{Adding noise \cite{chang2021defending}}&\multicolumn{2}{c}{DAP (proposed)}\\
     \cmidrule(lr){3-4} \cmidrule(lr){5-7} \cmidrule(lr){8-12} \cmidrule(lr){13-14}
    & &1*TERA&9*TERA&Median&Mean&Gaussian&$\sigma$=0.002&$\sigma$=0.005&$\sigma$=0.01&$\sigma$=0.02&$\sigma$=0.05&iter=1k&iter=80k\\
    \midrule
    genuine&0.828&20.890&35.818&27.853&1.161&10.973&1.161&1.547&2.277&3.675&9.731&2.505&1.253\\
    adv-PGD&91.683&48.356&42.360&27.557&88.935&26.722&23.188&5.609&3.727&4.762&9.524&\textbf{3.340}&7.087\\
    adv-BIM&92.133&26.499&39.545&26.087&76.789&13.872&6.576&2.901&2.484&3.868&9.731&2.321&\textbf{2.070}\\
    \bottomrule
    \end{tabular}}
    \vspace{-1.5em}
  \label{tab:defense}
\end{table*}

\subsubsection{ASV system}

We employed ECAPA-TDNN \cite{desplanques2020ecapa} as the main victim ASV model for adversarial attacks, which consists of convolutional layers with 512 channels. We used the AAM-Softmax objective function \cite{liu2019large} with hyperparameters \{s=32, m=0.2\} for training, along with attentive statistical pooling. The input acoustic feature is an 80-dimensional LogFBank representation with a 25ms hamming window and a 10ms step size. Additionally, cepstral mean and variance normalization (CMVN) is applied to the features. Data augmentation techniques including speed perturbing, superimposed disturbance, and reverberation enhancement are employed. Cosine distance is used to produce similarity scores between embeddings.

\subsubsection{Adversarial attack}

{We employed PGD attack \cite{ruff2021unifying} and BIM attack \cite{joshi2021study}, which adopt the same parameters \{$\epsilon$=30, $\alpha$=1\} to generate the adversarial examples. The iteration step was normally set as 50. To ensure a consistent signal-to-noise ratio (SNR) between genuine and adversarial examples, we added Gaussian white noise to the genuine examples, with the noise level determined by the corresponding adversarial perturbations. As a result, the mean signal-to-noise ratio for the genuine examples was set to approximately 40dB.}

\subsubsection{Evaluation metrics}

We measured the performance of the ASV systems and the effectiveness of the defense mechanism using two commonly used metrics: Equal Error Rate (EER) and minimum Detection Cost Function (minDCF) with $p$ = 0.01 and $C_{miss}$ = $C_{fa}$ = 1 \cite{nagrani2020voxsrc}. EER measures the point at which the false acceptance rate (FAR) equals the false rejection rate (FRR). minDCF is a cost-based metric that considers both false acceptance and false rejection errors, allowing for a more comprehensive evaluation of ASV system performance.

We evaluated the reconstruction performance of the purified signals using three objective metrics: Scale-Invariant Signal-to-Distortion Ratio (SI-SDR), Short-Time Objective Intelligibility (STOI), and Perceptual Evaluation of Speech Quality (PESQ). These metrics assess the quality and intelligibility of the audio, where higher values indicate better speech quality.

\subsubsection{Baseline and proposed methods}

For our Diffusion-Based Adversarial Purification (DAP) model, we adopted the same architecture as DiffWave \cite{kong2020diffwave}. To introduce controlled noise during the training stage, we set step $T$ as 100 and utilized a linear noise schedule \{0.0001, 0.035\}  to apply $\beta_t$ at each step. It makes $\beta_t$ begin with 0.0001 and increase to 0.035 over 100 steps. In the inference stage, the six-step variance schedule [0.0001, 0.001, 0.01, 0.05, 0.2, 0.35] is applied to set the value of $\gamma$ in the fast sampling algorithm. In addition, We used adversarial examples generated by the PGD method for training the DAP model. We trained two DAP systems conditioned on 512-dimensional spectrogram, which were trained by 1k and 80k iterations, respectively.

We compared the proposed adversarial defense method with three adversarial defense methods \cite{wu2021improving,wu2020defense,chang2021defending}. (i) The TERA method \cite{wu2021improving} employed a self-supervised model to reconstruct the acoustic feature. We pretrained it with the same setting in \cite{wu2021improving} except on 80-dim LogFBank features. (ii) The spatial smoothing method \cite{wu2020defense} used median, mean, and Gaussian filters to process the input audio. (iii) The noise-based method \cite{chang2021defending} added Gaussian noise to the entire audio signal. In our experiments, the standard deviation of the noise was set to \{0.002, 0.005, 0.01, 0.02, 0.05\}.

\subsection{Experimental results}

\begin{table}[t]
  \centering
  \caption{Quality of the audio signals that are first generated from PGD adversarial examples and then purified by defense models. }
  \vspace{-0.5em}
    \scalebox{1}{
    \begin{tabular}{ccccc}
    \toprule
    Defender&SI-SDR&STOI&WB-PESQ&NB-PESQ\\
    \midrule
    N/A&35.099&0.991&4.412&4.397\\
    \midrule
    Median filter \cite{wu2020defense}&-16.359&0.632&1.152&1.397\\
    Adding Noise \cite{chang2021defending}&\textbf{12.572}&0.867&1.556&2.353\\
    DAP(proposed)&11.467&\textbf{0.932}&\textbf{3.120}&\textbf{3.717}\\
    \bottomrule
    \end{tabular}
    }
  \label{tab:quality}
\vspace{-1em}
\end{table}

\begin{figure}[t]
\centering
\scalebox{1}{
\subfigure[Waveform]{
	\centering
	\label{fig:waveform}
	\includegraphics[width=0.22\textwidth]{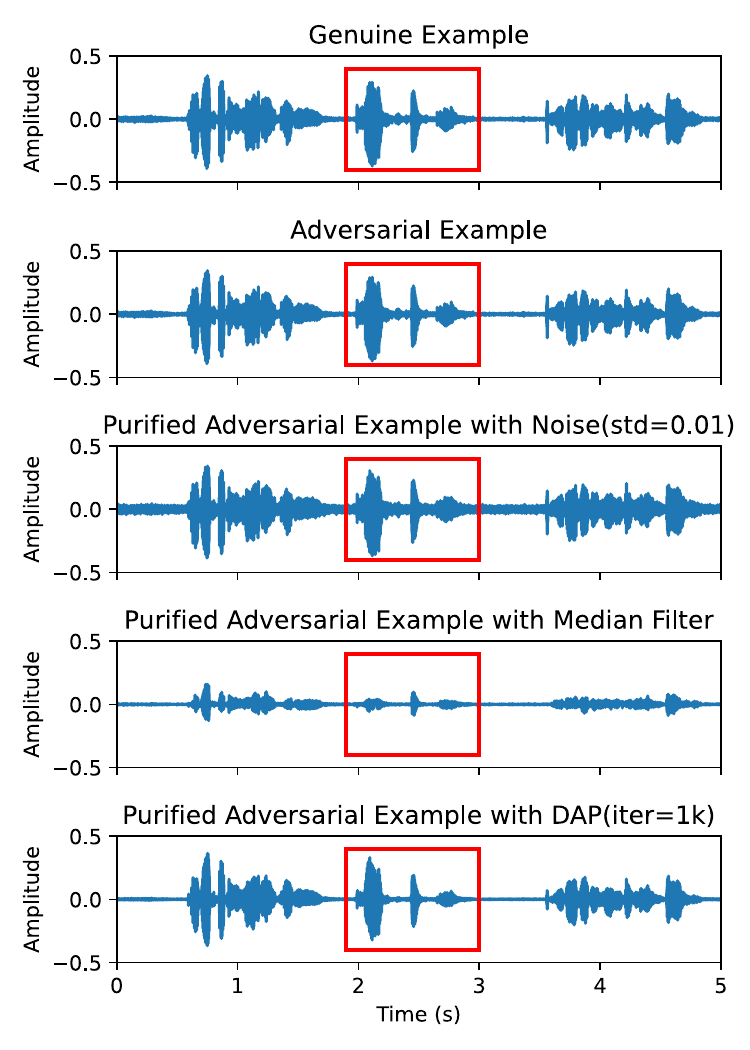}
}
\subfigure[Fbank features]{
	\centering
	\label{fig:spectrogram}
	\includegraphics[width=0.22\textwidth]{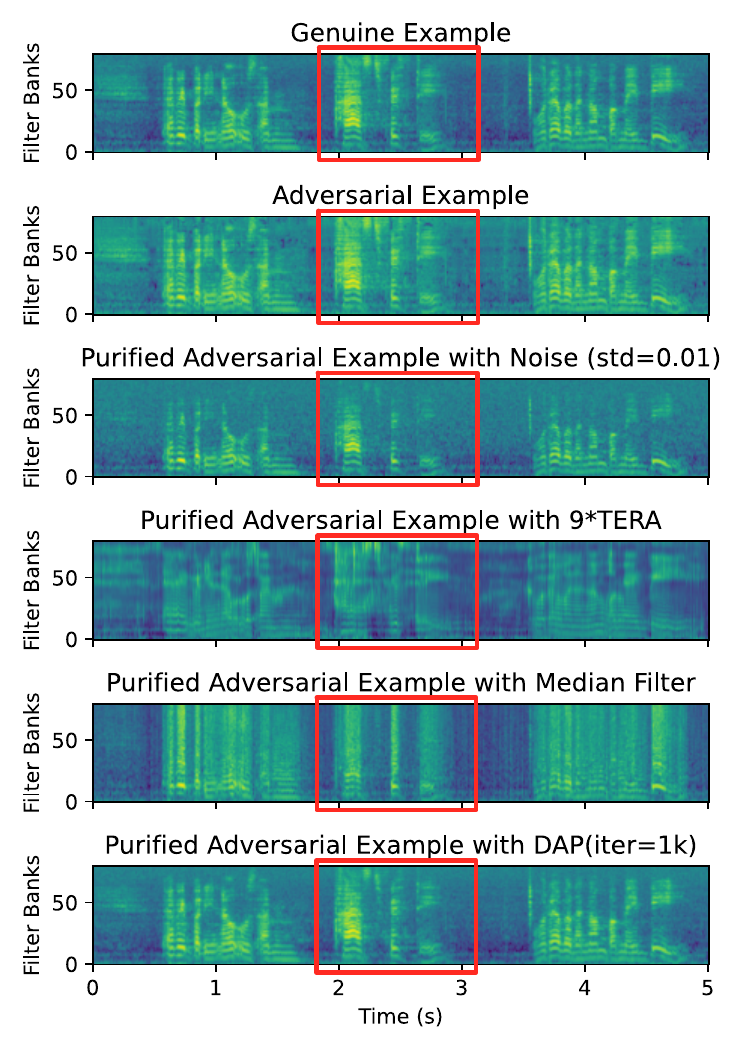}
}
}
\caption{
    A comparison example between the original audio and its adversarial example with different defenders. The genuine example is from id10270/5r0dWxy17C8/00024.wav of VoxCeleb1. As TERA method focuses on feature-level purification, it is not included in Fig. \ref{fig:waveform}.
}
\label{fig:wav}
\vspace{-1em}
\end{figure}

\begin{table}[t]
  \centering
  \caption{EER(\%) results of ECAPA-TDNN and Fast-ResNet34 under PGD and BIM attack methods.}
  \vspace{-0.5em}
    \scalebox{0.96}{
    \begin{tabular}{cccccccc}
    \toprule
    \multirow{2}[0]{*}{Attacker}&\multirow{2}[0]{*}{Steps}&\multicolumn{3}{c}{ECAPA-TDNN}&\multicolumn{3}{c}{Fast-ResNet34}\\
    \cmidrule(lr){3-5} \cmidrule(lr){6-8}
    & &N/A&Noise&DAP&N/A&Noise&DAP\\
    \midrule
    \multirow{4}[0]{*}{PGD}&10&83.232&3.340&3.758&85.339&2.714&2.128\\
    &20&95.633&3.704&4.115&96.378&3.132&2.277\\
    &50&98.668&4.983&4.293&98.992&3.833&2.899\\
    &100&98.998&5.010&4.528&99.398&3.967&2.901\\
    \midrule
    \multirow{4}[0]{*}{BIM}&10&59.872&3.311&3.520&62.128&2.505&1.852\\
    &20&84.258&3.448&3.868&87.628&2.714&2.244\\
    &50&94.821&3.643&4.348&95.899&3.132&2.996\\
    &100&95.834&3.883&4.348&96.604&3.292&3.125\\
    \bottomrule
    \end{tabular}
    }
    \vspace{-1.5em}
  \label{tab:attacker}
\end{table}

\subsubsection{Defense performance}

Table \ref{tab:defense} compares the DAP method with three purification methods on the ASV defense task. DAP outperforms others in defending against adversarial examples while maintaining performance on genuine examples. Furthermore, DAP is capable of defending against both $\ell_\infty$ and $\ell_2$ attacks. In addition, although self-supervised models like TERA are not trained for adversarial purification tasks, they could still demonstrate some defensive capabilities. We also evaluated another pretrained self-supervised model---WavLM \cite{chen2022wavlm}. It reaches an EER of 9.427\% after purifying the PGD examples.

\subsubsection{Reconstruction performance}

In terms of audio reconstruction quality, DAP surpasses other methods as shown in Table \ref{tab:quality}. Fig. \ref{fig:wav} gives a visualized comparison between the original audio and its corresponding adversarial examples after processed by different defense methods. From the figure, we see that the adversarial example purified by DAP are observed to be more similar to the original signal compared to those purified by the other approaches, e.g. the highlighted part in the red box. Moreover, DAP can reduce the noise and reverberation component of the speech signal. We also conducted a human listening test on 50 DAP-purified samples with five listeners. In 52\% of the cases, listeners could not discern any differences, while 36\% of the samples were perceived as similar to the original audio but with slight noise. For the remaining 12\%, listeners noticed significant noise.

\subsubsection{Effect of attack settings}

In this subsection, we study the robustness of the proposed method against different attackers and with different victim models. Table \ref{tab:attacker} lists the performance of the ECAPA-TDNN and Fast-ResNet34 victim models under the $\ell_2$ PGD and $\ell_\infty$ BIM attack methods, where the standard deviation of the adding noise method was set to 0.01. The results show that our method is effective under different attack scenarios and different ASV architectures.

\section{Conclusion}

In this letter, we propose a DAP method for the ASV defense against adversarial attacks. DAP utilizes a diffusion model to purify the adversarial examples and mitigate the perturbations in audio inputs. We conducted experiments in scenarios where the attackers are unaware of the defense method. The experimental results indicate that our approach outperforms the representative purification methods. It also introduces the minimal distortion to the genuine examples over the comparison methods.

\bibliographystyle{IEEEtran}
\normalem
\bibliography{ref}
\end{document}